# Magicarpet: A Parent-child Interactive Game Platform to Enhance Connectivity between Autistic Children and Their Parents

Yuqi Hu[1, *], Yujie Peng[1], Jennifer Gohumpu[1], Caijun Zhuang[2], Lushomo Malambo[1], Cuina Zhao[1]

1. Ningbo Innovation Center, Zhejiang University, China
2. NingboTech University, China

Autistic children often face challenges in social interaction and communication, impacting their social connectivity, especially with their parents. Despite the effectiveness of game-based interactive therapy in improving motor skills, research on enhancing parent-child relationships is lacking. We address this gap with Magicarpet, an interactive play carpet that encourages parent-child interaction and has been validated through a user study with five families. The preliminary results indicate that Magicarpet enhances the motivation and participation of autistic children in play, demonstrating the potential of human-computer interaction (HCI) designs to foster connectivity.

## Introduction

Autism Spectrum Disorder (ASD) is a neurodevelopmental disorder defined by core impairments in social interaction and communication, as well as restricted and/or repetitive behavior patterns [1]. Given these challenges, innovative approaches to therapy and rehabilitation are essential to support individuals with ASD in improving their quality of life. In recent years, exergames have been considered an effective way to motivate children to engage in rehabilitative exercises. These games can stimulate their interest and guide their attention by creating a relaxing, entertaining, and immersive rehabilitation atmosphere [2].

Parent-child interaction in the realm of physical activities and games has garnered attention, with a particular emphasis on the importance of increased play for stimulating exploratory activities that engage children more, especially within a social context [3]. Increased awareness among parents of the importance of health and interaction in autistic children increases sensitivity towards each other, thereby strengthening the parent-child bond [4]. Much research that focuses on game interaction encourages parental involvement in activities. Studies emphasized the game process should not only guide the child but also increase parents' participation and immersive experience [5]. Inclusive design is paramount in the development of such games, ensuring that they are accessible and accommodating to children with special needs, including those with ASD. The design should thoughtfully incorporate features that address the unique characteristics and requirements of this population. By doing so, the games can serve as a bridge, enhancing connectivity between autistic children and others, and endowing them with a form of "social capability" that may otherwise be challenging for them to develop naturally.

In this paper, we introduced Magicarpet, an interactive gaming platform that combines the theory of Reciprocal Imitation Training (RIT) [6] to improve activity engagement and dyadic synchrony, to strengthen the connection between parents and autistic children. It is designed to enhance the enthusiasm of autistic children in exergames by stimulating sensory experiences and increasing parental involvement. Magicarpet consists of two modes: 1) imitation mode: aiming at sequential synchronization through imitating parental movements; 2) collaboration mode: enhancing cooperative behaviors through collaborative tasks. A preliminary user study was conducted with five groups of families. The process of activities was recorded and semi-structured interviews were conducted with parents and therapists to understand the user experience with Magicarpet. Results indicated that Magicarpet was not only engaging but also effective in encouraging parents to participate enthusiastically. It fosters increased activeness and participation of children in exergames.

# Related work

## Exergame technologies in autistic research

Studies have shown that game-based activities accelerate the learning process in autistic children, enhancing attention and increasing the willingness to complete required tasks [7]. Incorporating games into daily education helps to encourage social interaction, develop communicative exchange and imaginative thinking, and enhance the various skills needed for children to carry out daily activities [8]. Among all, exergames exhibit the potential to support physical activity, foster motor development, and enhance the learning experience [9].

Games designed to enhance engagement and interaction for autistic children encompass both physical and digital ones. Physical interaction games, such as Mazi [10], a round and soft therapy ball-like device designed to promote collaboration and self-regulation during play, provide richer stimuli for autistic children. Olly[11] leverages haptic and auditory cues to improve socialization and participation. While, digital interactive games such as FutureGym [12], Lands of Fog [8], and Magic Room [13] all provide autistic children with interactive spaces and surfaces. Whole-body interaction can provide multi-sensory stimulation, and exercise-centered activities can keep autistic children focused [14]. One notable example within this domain is the Stomp game[15], which employs digital content displayed on the floor to create an interactive surface.

Physical interactive games can facilitate spontaneous interactive skills in autistic children through the stimulation of tangible objects and the feedback of digital augments. They hold the potential to improve motivation, cultivate social skills, heighten sensory awareness, and to some extent, encourage motor activities [7]. Physical games often require advanced cognitive and verbal skills, limiting accessibility for autistic children. Digital games encourage exercise but lack timely feedback, causing perceptual issues and high costs. Combining the strengths of both can overcome these limitations.

## Interactional synchrony and engagement/connectivity in autism research

Synchrony in activities is vital to engagement and interaction quality because it involves coordinated movement and communication between participants. Studies in HCI have highlighted the importance of both taking turns and acting in unison during interactions[16]. Notably, autism research has identified interactive synchronization—the natural alignment of movements during social activities — as essential for developing motor, social, and cognitive skills [17]. OSMoSIS [13] is a full-body interactive music system using motion capture to foster synchrony between autistic children and caregivers, enhancing social and communicative engagement. Synchrony levels indicate active participation and effective parent-child interactions.

To improve parent-child interaction synchronization, RIT can be used. RIT uses a blend of natural behavior and development strategies to teach imitation within a social-interactive context [17]. In practical terms, RIT might involve a series of interactive activities where the therapist or caregiver models an action, and the child is encouraged to imitate it. This could be as simple as clapping hands, waving goodbye, or more complex actions related to play or daily routines. The key is that the child is not just imitating for the sake of it but is doing so in a way that is meaningful within the context of social interaction [13]. Designs following the RIT method, like the interactive game carpet, can be adapted to help autistic children coordinate actions and participate in games, promoting interaction synchronization.

Research shows that Stomp games and synchronized parent-child interaction positively impact autistic children's motor skills training. Thus, an interactive carpet combining these elements with exergames has been developed, with similar designs successfully used for children with other special needs. An interactive rehabilitation carpet [18] has been developed for children with cerebral palsy, which increases interactivity with people by integrating engaging games, and visual, and auditory feedback. For children with Down Syndrome (DS), an interesting interactive mirror carpet [1] has been designed to support interesting interactions between parents and children and to support parent-child relationships. This interactive carpet exhibits the following advantages: 1) It can better adapt to the participation of multiple roles; 2) It has strong expandability. It not only can support coordinated training but also provides entertainment and a nurturing environment to promote social and intellectual development in children with special needs.

## Prototype design

### Hardware design

Based on the aforementioned studies, a prototype was developed. The mat consists of six layers (see Figure 1 left), from top to bottom: 1) Interchangeable Cover: This top layer could be changed to offer a variety of patterns and figures. 2) LED Panels Layer: This layer consists of two groups of a 3x2 array of LED panels. 3) Sensor Switch Layer Groups: Two copper layers separated by a sponge pad with holes. Pressure connects the copper layers, corresponding to one LED panel above. 4) Padding Layer: This bottom layer offers a comfortable touching and walking experience during use. There is an external circuit box to house the microprocessor and related circuit components. The hardware comprises: STM32-F103ZET6 microcontroller, 8x8x12 WS2812B LEDs, filter capacitors, and a patch switch. Programming in C modifies registers to change LED brightness, color, and quantity. Utilizing Analog-to-digital converter (ADC) sampling reads patch point levels to control LED on/off states.

### Gameplay mechanics

We designed two interaction modes for the mat: imitation and collaboration. Figure 1 (right) presents the two interaction modes of the mat. In imitation mode, parents light up a random light, and children imitate by hitting/stepping on the corresponding light on their side. In collaboration mode, the system lights up several random lights, and both parents and children work together to turn them off. Flashes prompt actions if none are detected within a certain interval. These modes describe the interaction mechanism but do not limit user interaction, allowing for open-ended play and the development of unique ways of playing. However, the "imitation" and "collaboration" are only descriptions of the interaction mechanism, it does not limit the way how users interact with it. Thus, they can still have open-ended play and develop their ways of playing.

Table 1 shows the design elements of the Magicarpet, aligning each feature with the theoretical principles of RIT to enhance activity engagement and dyadic synchrony in parent-child interactive exergames.

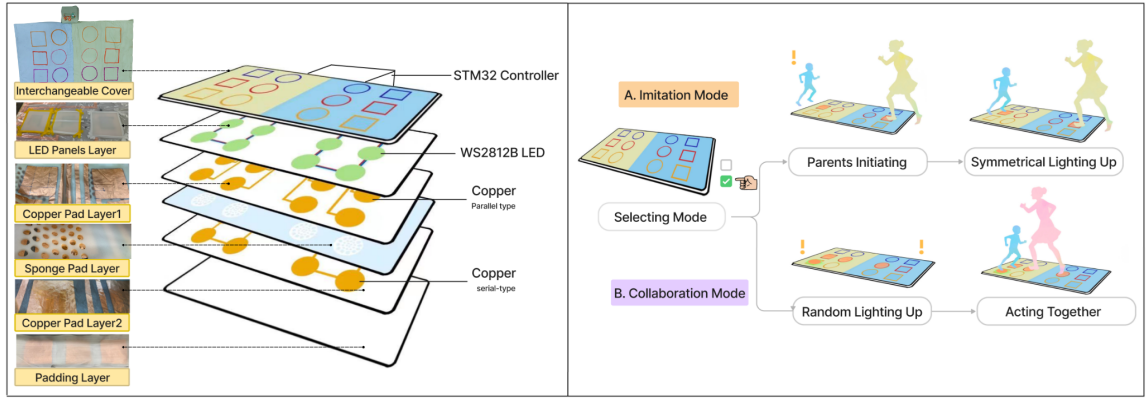

Figure 1: Layout structure of the Magicarpet (left), and two play modes of the Magicarpet (right).

Table 1: Description of the design element of the Magicarpet according to RIT's theoretical framework

| Design Element | Using RIT for Effective Auxiliary Intervention |
| --- | --- |
| Color Partition | **Visual Structuring:** RIT benefits from a structured learning environment with color coding that organizes the space and clarifies roles and actions for children. |
| Button Design | **Visual Cues:** The variation in color and shape provides visual guidance for children to identify and navigate specific movement areas in RIT. |
| Light Feedback | **Clear Visual Cues:** RIT uses clear visual cues, like lights, to provide immediate feedback, helping children understand the consequences of their actions and learn through imitation. |
| Imitation Mode | **Imitation Learning:** At the core of RIT is imitation learning, where children learn to extinguish lights by observing and replicating their caregivers' actions. |
| Collaboration Mode | **Shared Experience:** RIT involves shared experiences, like extinguishing lights together, which strengthen the parent-child bond. |

## User study

Our team conducted user research at one integrated education institution in Zhuji City, Zhejiang Province, China. The goals of our user research were: 1) determine if children and their parents could easily understand, enjoy, and use the carpet; 2) evaluate whether the game modes helped with improved connectivity in parent-child interaction.

Table 2: Child Participant Information

| Participants | P1 | P2 | P3 | P4 | P5 |
| --- | --- | --- | --- | --- | --- |
| Age | 6 | 5 | 5 | 6 | 7 |
| Autism Level | Level-2 | Level-1 | Level-2 | Level-2 | Level-3 |

### Participants

We recruited five groups of participants, consisting of four boys and one girl, aged 5-7 years old (M age=5.8 years). All of these children had been diagnosed with ASD and were undergoing long-term, full-day interventions. The autism level of all children was assessed according to the fifth edition of the Diagnostic and Statistical Manual of Mental Disorders (DSM-5) [19]. Table 2 shows the detailed information of each child. The biological mothers of each child participated in the study.

### Procedure

The experiment took place in a classroom with daily interventions. Two cameras recorded the parent-child play process. Five parent-child groups signed informed consent forms and provided demographic information, including their daily training details. Figures 2 and 3 show experiment photos and procedures. Each family played both game modes in random order, with each activity lasting 10-15 minutes based on the child's performance. Semi-structured interviews with parents followed the game sessions, discussing the child's performance and providing feedback, rated on a Likert scale. Additionally, five therapists were interviewed to assess the performance and give their opinions and suggestions.

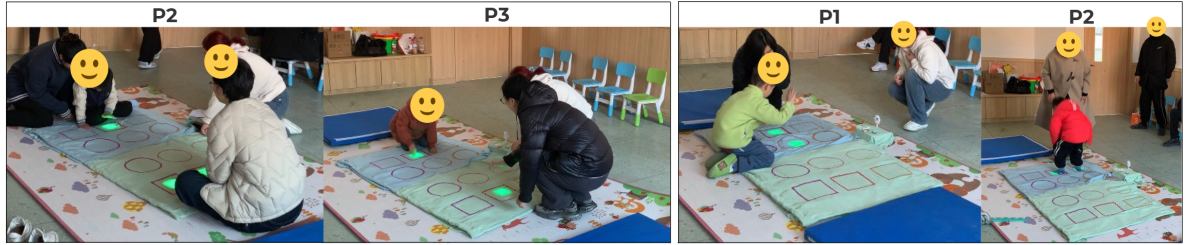

Figure 2: Photos taken during the user study: "Imitation mode" (left) and "Collaboration mode" (right).

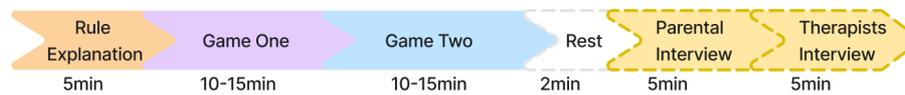

Figure 3: Procedure of the User Study.

## Data collection and analysis

We collected datasets using three methods: questionnaires, semi-structured interviews, and video recording. We used the Human Observation Coding (HOC) [20] to define two movement types for children's responses to imitation instructions in the imitate modes. HOC facilitates the quantification of children's engagement levels during interactive activities by systematically documenting and categorizing their behavioral patterns, such as active participation, passivity, or non-participation. Empirical observations suggest that frequent and precise imitative actions are indicative of a high degree of engagement. Child's total HOC score ($HOC_c$) was calculated by averaging this child's normalized movement type scores. Note that repetition items were not considered as elements of a movement type and hence were not counted towards the variable $k_m$. The equation used to calculate the children's overall HOC score was as follows [20]:

$$HOC_c = \frac{1}{M}\sum_{m=1}^{M} s_{cm} = \frac{1}{M}\sum_{m=1}^{M} \frac{s_{pos_{cm}} + s_{neg_{cm}}}{k_m} \quad (eq.1)$$

Two experimenters watched the videos to identify the target behaviors in intimation mode according to the coding scheme based on the HOC. Positive items ($s_{pos}$) include completed (score:1) and not completed (score:0) behaviors. Negative items ($s_{neg}$) include not performing on the reverse side or repeating more than demonstrated (score:0), performing on the reverse side (score: -0.5), and correctly performing but repeating more times than demonstrated (score: -1). In the collaboration mode, experimenters marked the number of times that the child and parent successfully worked together to eliminate light spots, as the completion degree in this activity.

## Findings and result

### Game experience: Enjoyment and participation

The results of the semi-structured interviews with parents suggest that compared to regular gross motor activities, both the imitation and collaboration activities in the parent-child interactive carpet were very appealing to the children, significantly enhancing their enthusiasm and participation in gross motor skill training activities.

As shown in Table 3, in the imitation mode questionnaire, three children were able to coordinate well with their parents, understanding and following their parents' instructions, with increased levels of engagement and concentration. In the collaborative mode questionnaire, children's completion rates and emotional responses were higher than in the imitation games, as parents unconsciously assisted the children in achieving game objectives during collaboration, and the children felt excited by their parents' help.

Table 3：Imitation Mode and Collaborative Mode Questionnaire Survey Results

| Question (Compared to the routine training) | Imitation Mode | | | | | | Collaboration Mode | | | | | |
|---|---|---|---|---|---|---|---|---|---|---|---|---|
| | P1 | P2 | P3 | P4 | P5 | Avg | P1 | P2 | P3 | P4 | P5 | Avg |
| I. What is your overall impression of our system? (1: negative – 5: positive) | 5 | 5 | 4 | 3 | 2 | 3.8 | 5 | 5 | 4 | 3 | 2 | 3.8 |
| II. How would you rate the child's engagement level? (1: low – 5: high) | 5 | 5 | 4 | 3 | 2 | 3.8 | 5 | 5 | 4 | 3 | 2 | 3.8 |
| III. How would you rate the child's attentiveness level? (1: low – 5: high) | 5 | 5 | 3 | 3 | 2 | 3.6 | 5 | 5 | 3 | 3 | 2 | 3.6 |
| IV. How would you rate the child's activity completion level? (1: low – 5: high) | 4 | 5 | 3 | 3 | 2 | 3.4 | 5 | 5 | 4 | 3 | 2 | 3.8 |
| V. How would you rate the child's emotional level? (1: low – 5: high) | 5 | 5 | 4 | 2 | 2 | 3.6 | 5 | 5 | 4 | 3 | 2 | 3.8 |
| VI. How would you rate the child's understanding of the game's rules? (1: easy – 5: hard) | 5 | 5 | 3 | 3 | 2 | 3.6 | 5 | 5 | 3 | 3 | 2 | 3.6 |
| Avg | 4.8 | 5 | 3.5 | 2.83 | 2 | | 5 | 5 | 3.6 | 3 | 2 | |

During the activity, the therapists noted that "P1 and P2 displayed more activeness than usual, purposefully and autonomously learning the parent's tapping gestures to react more swiftly." P5, who had severe autism (Level-3) lacked patience during the exercises, and the experiment, and exhibited challenging behaviors such as aggression, non-compliance, and other stereotyped actions. Such challenging behaviors are common among children with autism in this age group, and severely challenging behaviors may affect the learning opportunities for autistic children to acquire skills [18]. Therefore, we infer that the current design might not be suitable for children with severe autism. During training, parents' suggestive language increased children's enthusiasm and interest in the carpet activities. For example, P1's parent enhanced the imitation game by creating a picnic story, turning purple circles into "grapes" and encouraging the child to "eat" them. In the collaboration game, P1 imagined lit-up lights as candles to be "blown out." This inspires us to incorporate scenarios into the game, rewarding children after specific actions. P2 described the activity as "catching mice," and P4 carried the child on their back, enhancing intimate interaction.

## Parent-children Interaction: Intimacy and Synchronicity

In order to evaluate children's activity completion in the process of interactive games, we investigated $HOC_c$ in imitation mode and the amount of task success in the cooperation mode. Completion degree refers to children and parents have successfully worked together to eliminate points in the cooperation mode. In general, the severity of autism is inversely proportional to the $HOC_c$, with higher severity corresponding to smaller $HOC_c$. $HOC_c$ between 0.4 and 0.7 indicates normal imitation performance.

For all cases, $HOC_c$ greater than 0.7 in autistic children indicates better imitation performance and relatively good motor imitation abilities. On the other hand, $HOC_c$ less than 0.4 indicates poorer imitation performance, leading to lower performance coefficients [20]. As shown in Table 4, all children's $HOC_c$ values are within the normal high range. Moreover, the $HOC_c$ values of P1 and P2 exceed the standard level, indicating a significant improvement in their parent-child interaction synchronization during the activity. Additionally, the completion degree values of P1 and P2 are also higher than the average, indicating their higher enthusiasm and participation compared to other children in the activity.

Table 4: The results of behavior coding

| Result | P1 | P2 | P3 | P4 | P5 | Avg |
|---|---|---|---|---|---|---|
| $HOC_c$ | 0.72 | 0.86 | 0.66 | 0.68 | 0.45 | 0.67 |
| Completion Degree | 8 | 10 | 5 | 6 | 1 | 6 |

During the interviews with parents and therapists, most parents believed that the interactive carpet did not limit their children's creativity in their play interactions. This means that the carpet can be used for gross motor coordination training, as well as for children's exploration of lights and crawling abilities. At the same time, this also makes parents more willing to explore the possibilities of the carpet with their children, as the carpet itself is filled with interesting interactions and a sense of intimacy. Five therapists reported that "unlike passive participation in traditional therapy, children were particularly active and proactive when using the interactive carpet for gross motor activities, due to the involvement of parents." P2's parents were satisfied with P2's performance, and said "he was particularly active on the carpet, he was lying down and rolling across the carpet to turn off lights at my encouragement." While P3 took the initiative to ride on his mother's back, asking her to complete tasks with him.

During the process of parent-child interaction, children gave positive feedback on the task invitations from their parents. P1 stated that "the combination of the carpet with engaging interactions enhanced the shared interests between the child and parent, and increased the synchronicity of our interactions."

# Discussion

This study exhibited certain limitations. In terms of game design, the differences in abilities between children of different levels were not considered, resulting in the activity being unattractive to children with severe autism and lacking challenge for children with high functional disabilities. Concerning experimental design, the sample size was small with a gender bias, and long-term comparative experiments were needed.

Interviews with therapists and participants highlighted the need for diverse games and feedback mechanisms, graded training programs tailored to varying abilities, and the integration of interactive carpets with curriculum content. Suggestions also included enhancing product sensitivity for children with weaker upper limb ability. Collecting more quantitative data, behavior labels, and physiological signals will improve the measurement of task load and system usability in future work.

# Conclusion

In this article, we proposed Magicarpet, a parent-child interactive platform designed to assist autistic children in participating actively in gross motor training. This research led to the development of two interactive modes: imitation and collaboration, targeting parent-child interaction synchronization and social communication skills, respectively. Preliminary results indicated that Magicarpet enhances autistic children's experience in gross motor coordination activities and improves parent-child interaction synchronization. This research emphasizes the significance of specialized parent-child interactive games in promoting engagement in gross motor activities and has contributed to the design and HCI field for autistic children. By involving parents as participants, children's enthusiasm is stimulated by shared participation rather than one-way instructions. The scalability of the interactive carpet serves as an engaging learning and social tool for autistic children in exergames. Further research is needed to verify the effectiveness of the two game modes and explore their long-term use.

## Acknowledgments

We extend our heartfelt gratitude to the dedicated teachers and the families who participated in the study, for their generous contribution and invaluable collaboration.

## References

[1] Stefan Manojlovic, Laurens Boer, and Paula Sterkenburg. 2016. Playful interactive mirroring to support bonding between parents and children with Down Syndrome. In Proceedings of the The 15th International Conference on Interaction Design and Children, June 21, 2016. ACM, Manchester United Kingdom, 548–553. https://doi.org/10.1145/2930674.2935987

[2] Karina Caro. 2014. Exergames for children with motor skills problems. SIGACCESS Access. Comput. 108 (January 2014), 20–26. https://doi.org/10.1145/2591357.2591360

[3] Guilherme Azevedo, Mateus Ludwig Gunsch, Mariane Gomes Lacerda, Henrique De Oliveira Peixoto, Luciana Correia L. F. Borges, and Eunice P. Dos Santos Nunes. 2023. Requirements Gathering Regarding Fine Motor Skills, Adaptive Difficulty and Executive Functions for a Game in Development for Therapy Sessions with Autistic Children to Encourage Collaboration. In Proceedings of the 2023 ACM International Conference on Interactive Media Experiences Workshops, June 12, 2023. ACM, Nantes France, 32–37. https://doi.org/10.1145/3604321.3604345

[4] Taeahn Kwon, Minkyung Jeong, Eon-Suk Ko, and Youngki Lee. 2022. Captivate! Contextual Language Guidance for Parent–Child Interaction. In CHI Conference on Human Factors in Computing Systems, April 29, 2022. ACM, New Orleans LA USA, 1–17. https://doi.org/10.1145/3491102.3501865

[5] Brooke Ingersoll. 2012. Brief Report: Effect of a Focused Imitation Intervention on Social Functioning in Children with Autism. J Autism Dev Disord 42, 8 (August 2012), 1768–1773. https://doi.org/10.1007/s10803-011-1423-6

[6] Jiayu Lin, Jiefeng Li, Jielin Liu, Yingying She, Fang Liu, Zhu Zhu, Mingbo Hou, Lin Lin, and Hang Wu. 2023. Supporting Autistic Children's Group Learning in Picture Book Reading Activity with a Social Robot. In Proceedings of the 22nd Annual ACM Interaction Design and Children Conference, June 19, 2023. ACM, Chicago IL USA, 480–485. https://doi.org/10.1145/3585088.3593875

[7] Laura Bartoli, Clara Corradi, Franca Garzotto, and Matteo Valoriani. 2013. Exploring motion-based touchless games for autistic children's learning. In Proceedings of the 12th International Conference on Interaction Design and Children, June 24, 2013. ACM, New York New York USA, 102–111. https://doi.org/10.1145/2485760.2485774

[8] Joan Mora-Guiard, Ciera Crowell, Narcis Pares, and Pamela Heaton. Sparking social initiation behaviors in children with Autism through full-body Interaction.

[9] Karina Caro, Ana I. Martínez-García, Mónica Tentori, and Iván Zavala-Ibarra. 2014. Designing exergames combining the use of fine and gross motor exercises to support self-care activities. In Proceedings of the 16th international ACM SIGACCESS conference on Computers & accessibility - ASSETS '14, 2014. ACM Press, Rochester, New York, USA, 247–248. https://doi.org/10.1145/2661334.2661403

[10] Bernd Huber, Richard F. Davis, Allison Cotter, Emily Junkin, Mindy Yard, Stuart Shieber, Elizabeth Brestan-Knight, and Krzysztof Z. Gajos. 2019. SpecialTime: Automatically Detecting Dialogue Acts from Speech to Support Parent-Child Interaction Therapy. In Proceedings of the 13th EAI International Conference on Pervasive Computing Technologies for Healthcare, May 20, 2019. ACM, Trento Italy, 139–148. https://doi.org/10.1145/3329189.3329203

[11] Maryam Jahadakbar, Carlos Henrique Araujo De Aguiar, Arman Nikkhah Dehnavi, and Mona Ghandi. 2023. Sounds of Play: Designing Augmented Toys for Children with Autism. In Proceedings of the 16th International Conference on PErvasive Technologies Related to Assistive Environments, July 05, 2023. ACM, Corfu Greece, 338–346. https://doi.org/10.1145/3594806.3594859

[12] Issey Takahashi, Mika Oki, Baptiste Bourreau, Itaru Kitahara, and Kenji Suzuki. 2018. FUTUREGYM: A gymnasium with interactive floor projection for children with special needs. International Journal of Child-Computer Interaction 15, (March 2018), 37–47. https://doi.org/10.1016/j.ijcci.2017.12.002

[13] Grazia Ragone, Kate Howland, and Emeline Brulé. 2022. Evaluating Interactional Synchrony in Full-Body Interaction with Autistic Children. In Interaction Design and Children, June 27, 2022. ACM, Braga Portugal, 1–12. https://doi.org/10.1145/3501712.3529729

[14] Matthew Ford, Peta Wyeth, and Daniel Johnson. 2012. Self-determination theory as applied to the design of a software learning system using whole-body controls. In Proceedings of the 24th Australian Computer-Human Interaction Conference, November 26, 2012. ACM, Melbourne Australia, 146–149. https://doi.org/10.1145/2414536.2414562

[15] Mathilde Bekker. 2007. Stimulating Children's Physical Play through Interactive Games: Two Exploratory Case Studies. (2007).

[16] Casey J. Zampella, Evangelos Sariyanidi, Anne G. Hutchinson, G. Keith Bartley, Robert T. Schultz, and Birkan Tunç. 2021. Computational Measurement of Motor Imitation and Imitative Learning Differences in Autism Spectrum Disorder. In Companion Publication of the 2021 International Conference on Multimodal Interaction, October 18, 2021. ACM, Montreal QC Canada, 362–370. https://doi.org/10.1145/3461615.3485426

[17] Brooke Ingersoll. 2010. Brief Report: Pilot Randomized Controlled Trial of Reciprocal Imitation Training for Teaching Elicited and Spontaneous Imitation to Children with Autism. J Autism Dev Disord 40, 9 (September 2010), 1154–1160. https://doi.org/10.1007/s10803-010-0966-2

[18] Yijia An, Qinglei Bu, Jie Sun, Eng Gee Lim, Lijun Kong, Zhiqin Chen, and Roshan Devaraj. 2023. Interactive Rehabilitation Carpet for Children with Cerebral Palsy. In Proceedings of the Seventeenth International Conference on Tangible, Embedded, and Embodied Interaction, February 26, 2023. ACM, Warsaw Poland, 1–5. https://doi.org/10.1145/3569009.3573110

[19] Diagnostic and Statistical Manual of Mental Disorders (DSM–5) [Internet]. Arlington, VA: American Psychiatric Assocation. About DSM-5 and Development; [cited 2017 May 08]. Available from: https://www.psychiatry.org/psychiatrists/practice/dsm/about-dsm."

[20] Tunçgenç B, Pacheco C, Rochowiak R, Nicholas R, Rengarajan S, Zou E, Messenger B, Vidal R, Mostofsky SH. Computerized Assessment of Motor Imitation as a Scalable Method for Distinguishing Children With Autism. Biol Psychiatry Cogn Neurosci Neuroimaging. 2021 Mar;6(3):321-328. doi: 10.1016/j.bpsc.2020.09.001.